\documentclass[a4paper,12pt,dvipsnames,usernames]{article}
\pdfoutput=1

\usepackage{aas_macros}

\usepackage{jcappub} 

\usepackage[utf8]{inputenc}
\usepackage{graphicx}
\usepackage{amsmath}
\usepackage{amssymb}
\usepackage{bm}
\usepackage{paralist,array}
\usepackage{units} 
\graphicspath{{./Plots/}}
\usepackage{slashed}
\usepackage{todonotes}
\usepackage[parfill]{parskip}
\usepackage{longtable}
\usepackage{float}
\usepackage{pdfpages}
\usepackage{gensymb}

\usepackage{caption}
\usepackage{subcaption}

\graphicspath{{Plots/}}

\title{Dark photon constraints from CMB temperature anisotropies}

\author[1]{Andres Aramburo-Garcia,}
 \emailAdd{aramburo@lorentz.leidenuniv.nl}
\affiliation[1]{%
 Institute Lorentz, Leiden University, Niels Bohrweg 2, Leiden, NL-2333 CA, the Netherlands
}

\author[2]{Kyrylo Bondarenko,}
 \emailAdd{kyrylo.bondarenko@su.se}
\affiliation[2]{
Nordita, KTH Royal Institute of Technology and Stockholm University, Hannes Alfv\'ens v\"ag
12, 10691 Stockholm, Sweden}

\author[1]{Alexey Boyarsky,}
 \emailAdd{boyarsky@lorentz.leidenuniv.nl}

\author[3]{Pavlo Kashko,}
 \emailAdd{pavlo.kashko@vub.be}
\affiliation[3]{Inter-University Institute for High Energies (IIHE), Vrije Universiteit Brussel, Brussels, 1050, Belgium}

\author[4,5]{Josef Pradler,}
\emailAdd{josef.pradler@univie.ac.at}
\affiliation[4]{Institute of High Energy Physics, Austrian Academy of Sciences, Georg-Coch-Platz 2, 1010 Vienna, Austria}
\affiliation[5]{University of Vienna, Faculty of Physics, Boltzmanngasse 5, A-1090 Vienna, Austria}

\author[6,7]{Anastasia Sokolenko,}
\emailAdd{sokolenko@kicp.uchicago.edu}
\affiliation[6]{
Theoretical Astrophysics Department, Fermi National Accelerator Laboratory, Batavia, Illinois, 60510, USA
}
\affiliation[7]{
Kavli Institute for Cosmological Physics, The University of Chicago, Chicago, IL 60637, USA
}

\author[8]{Roi Kugel,}

\author[1,8]{Matthieu Schaller,}
\affiliation[8]{Leiden Observatory, Leiden University, PO Box 9513, NL-2300 RA Leiden, The Netherlands}

\author[8]{Joop Schaye}

\begin{document}
\vspace*{-1.5cm}
\abstract{
The resonant conversion, within the inter-galactic medium, of regular photons into dark photons amplifies the anisotropy observed in the CMB, thereby imposing stringent constraints on the existence of light dark photons. In this study, we investigate the impact of light dark photons, with masses in the range $3\times 10^{-15} ~\rm{eV} < m_{A'} < 3\times 10^{-12}~\rm{eV}$ on the power spectrum of temperature anisotropies within the cosmic microwave background (CMB) radiation utilizing the state-of-the-art large-volume FLAMINGO cosmological simulations. Our results show that using full Planck data, one can expect the existing constraints on the dark photon mixing parameter in this mass range to improve by an order of magnitude.
}

\maketitle

\section{Introduction}
\label{sec:Introduction}

Dark photons are theoretical vector particles that interact with the Standard Model (SM) through kinetic mixing with the ordinary photon~\cite{Okun:1982xi,Galison:1983pa,Holdom:1985ag}. A method to search for dark photons involves observing the conversion of regular photons into dark ones. This conversion process is generally characterized by a low probability, suppressed by the small vacuum mixing angle between the dark and ordinary photons. However, in specific instances, when the effective, in-medium photon mass equals the dark photon rest mass, the probability of conversion can be significantly amplified, a phenomenon known as resonant conversion~(see, e.g., \cite{Georgi:1983sy, Jaeckel:2008fi,Mirizzi:2009iz}).

This phenomenon can be detected by analyzing the temperature anisotropies and the blackbody spectral distortions of the Cosmic Microwave Background (CMB). Specifically, the resonant conversion of CMB photons into dark photons manifests itself as a ``disappearance'' of CMB photons. The likelihood of this conversion is contingent on the photon's energy and varies with the observation direction, leading to CMB spectral distortions and sourcing additional temperature anisotropies. In previous studies~\cite{Jaeckel:2008fi,Mirizzi:2009iz,Caputo:2020bdy,Bondarenko:2020moh,Garcia:2020qrp,Caputo:2020rnx,Acharya:2022vck}, the limits on the dark photon mixing angle were derived from using constraints on CMB distortions from COBE/FIRAS~\cite{Fixsen:1996nj} and on the total variance of the temperature anisotropies from Planck~\cite{Planck:2018bsf} and SPT~\cite{George:2014oba}.

In this work, we improve on previous constraints on the dark photon mixing angle by using CMB temperature power spectrum measurements. The resonant conversion into dark photons post-reionization introduces additional late-time temperature anisotropies. They add to the respective power spectrum at large $\ell$ (small angular scales) as they correlate with the Universe's large-scale structure. For large $\ell$, the measured CMB power spectrum is comparatively suppressed, allowing us to derive stronger constraints on the amplitude of generated anisotropies. This approach should give stronger constraints than from the total variance, i.e.~$\ell$-summed power spectrum, as was done in prior work~\cite{Bondarenko:2020moh, Garcia:2020qrp}.

Specifically, we focus on resonant conversion after redshift $z=6$. At higher redshift, we expect additional contributions to the conversion probability during inhomogeneous reionization, but our current knowledge of this transition is limited. Additionally, there is a contribution to the conversion probability from even higher redshifts, before reionization, but the Universe was much more homogeneous then, and we do not expect a significant contribution to anisotropic conversion from this early time. Because we exclude the high-redshift contribution to the conversion probability, our results are conservative (see e.g.~\cite{Garcia:2020qrp} for discussion).

The structure of the paper is as follows: in Sec.~\ref{sec:Dark_photon_resonant_conversion}, we outline the physics underlying resonant conversion. Sec.~\ref{sec:simulations} introduces the FLAMINGO simulation suite~\cite{Schaye:2023jqv} and its application in generating a map of the conversion probability for a localized region on the sky. In Sec.~\ref{sec:results}, we establish constraints on the dark photon model by contrasting the temperature anisotropy power spectrum predicted by simulations with observed data. We present our conclusions in Sec.~\ref{sec:conclusions}.

\section{Dark photon resonant conversion}
\label{sec:Dark_photon_resonant_conversion}
The Lagrangian of the photon-dark photon system can be written as
\begin{equation}
\label{L}
    \mathcal{L}_{A+A'} =  -\frac{1}{4} F_{\mu \nu}^2 + e A_\mu J_{\rm em}^\mu - \frac{1}{4} (F_{\mu \nu}')^2 + \frac{1}{2} m_{A'}^2 (A'_{\mu})^2 - \frac{\epsilon}{2} F_{\mu \nu} F^{' \mu \nu},
\end{equation}
where the first two terms are for the SM photon $A_{\mu}$ with its coupling to the  electromagnetic current $J_{\rm em}^{\mu}$. The subsequent two terms are for the dark photon $A'_\mu$ with mass $m_{A'}$, and the final term gives the kinetic mixing between  $A$ and $A'$ with their respective field strength tensors $F_{\mu\nu}$ and $F'_{\mu\nu}$.%
\footnote{The origin of $m_{A'}$ is not relevant for this work; it can be induced by the spontaneous symmetry breaking of the $U(1)'$ or be of St\"uckelberg type.} The coupling $\epsilon$ enables the conversion process between the two states (see e.g.~\cite{Mirizzi:2009iz}).

In this work, we are concerned with photon-dark photon conversion in the intergalactic medium. Because of the interaction between a photon and particles in the intergalactic medium, the vacuum photon dispersion relation $ \omega^2 - \bm{k}^2 = 0$  is modified to $\omega^2 - \bm{k}^2 = m_{A}^2$.  Here, $\omega$ is the photon energy, $ \bm{k}$ the photon three-momentum, and $m_A$ is the effective photon mass. This changes the mixing angle between photon and dark photon from its vacuum value $\epsilon$ to (see e.g.~\cite{An:2013yfc}),
\begin{align}
\epsilon_{\rm eff}^2 = \frac{\epsilon^2 m_{A'}^4}{(m_{A}^2 - m_{A'}^2)^2}.
\end{align}
In the intergalactic medium after reionization, and for photon energies in the infrared band, the dominant contribution to the photon effective mass comes from the interaction with free electrons~\cite{Mirizzi:2009iz,Garcia:2020qrp,Caputo:2020bdy,Caputo:2020rnx} and it is given by the plasma frequency,
\begin{equation}
     m_{A}(n_e) \approx \sqrt{\frac{4\pi \alpha_{\text{em}} n_e}{m_e}},
     \label{eq:mA}
\end{equation}
where $n_e$ is the free electron number density, $\alpha_{\text{em}} \simeq 1/137$ is the electromagnetic fine-structure constant, and $m_e$ is the electron mass.

When the mass-squared of the effective photon equals that of the dark photon, $m_A^2 = m_{A'}^2$, the mixing angle between these states reaches its maximum value leading to resonant conversion, see~\cite{Mirizzi:2009iz}. The conversion probability is expressed as $P_{A' \to A}(\omega) = 1 - p$, where $p$ represents the level-crossing probability within the two-state system~\cite{Mirizzi:2009iz}. The latter is given by $p = \exp[- \pi \epsilon^2 m_{A'}^2 R/\omega]$, with $\omega$ denoting the photon energy, and the $R$-factor, a measure of the rate of change of the effective photon mass with time, is given by~\cite{Mirizzi:2009iz}
\begin{equation}
    R = \left|\frac{d \ln m_A^2}{d t}\right|^{-1}_{t = t_{\text{conv}}},
    \label{eq:Rfactor}
\end{equation}
Here, $t_{\text{conv}}$ represents the time at which the resonance occurs, satisfying $m_{A}^2(t_{\text{conv}}) = m_{A'}^2$. The conversion is considered adiabatic when the effective photon mass changes slowly with time, resulting in a conversion probability nearing unity. Conversely, in the non-adiabatic regime, the conversion probability is approximated by
\begin{equation}
    P_{A' \to A} = P_{A \to A'} \simeq \epsilon^2 \frac{\pi m_{A'}^2}{\omega} R.
    \label{eq:P-conv-Josef}
\end{equation}
Using the full expression for the transition probability is important when $P_{A' \to A}$ approaches unity~\cite{Garcia:2020qrp}. In this study, we restrict our analysis to the regime where $P_{A' \to A} \ll 1$ and the approximation above suffices. Given that the effective photon mass is solely a function of the electron number density, we define the resonant electron number density ($n_e^{\text{res}}$) through the condition $m_A^2(n_e^{\text{res}}) = m_{A'}^2$. Consequently, resonant conversion occurs exclusively at points where the electron number density matches~$n_e^{\text{res}}$.

When a photon traverses the Universe, it may satisfy the resonant conversion condition multiple times. Consequently, we can define the total probability of resonant conversion, $P_{\text{tot}}$, as the cumulative probability of photon conversion along its trajectory (see, for example,~\cite{Garcia:2020qrp})
\begin{align}
     P_{\text{tot}}(\omega) \approx \sum_i P_i =
    \frac{\pi \epsilon^2 m_{A'}^2}{\omega} \sum_i \frac{R_i}{(1+z_{i})} \theta(z_{\rm max} - z_i),
    \label{eq:ptot}
\end{align}
where $P_i$ denotes the conversion probabilities of individual resonances at their respective redshifts $z_i$ with the corresponding $R$-factor $R_i$, under the assumption that $P_{\text{tot}} \ll 1$ holds. Function $\theta$ above is a Heaviside step function.

\section{Simulations}
\label{sec:simulations}

\subsection{Description of the simulations}

The FLAMINGO\footnote{\url{https://flamingo.strw.leidenuniv.nl/}} suite~\cite{Schaye:2023jqv,Kugel2023} comprises a collection of hydrodynamical and dark matter only simulations executed using \textsc{swift}~\cite{Swift:2023}, an open-source, cosmology,  hydrodynamics and gravity solver code. The suite includes 16 hydrodynamic and 12 dark matter-only simulations~\cite{Schaye:2023jqv} and investigates various resolutions, galaxy formation, and cosmological models. It comprises the largest hydrodynamical simulation run to~$z = 0$ to this day. In our study, we employ the hydrodynamical simulation volume with a box side length of $L = 1~\mathrm{cGpc}$ (comoving Gpc) and adopting the suite's fiducial cosmology and galaxy formation models; the cosmological parameters are taken from the maximum-likelihood values of the Dark Energy Survey~\cite{DESY_cosmoparm} including external constraints. The subgrid model for feedback from star formation and active galactic nuclei was calibrated using machine-learning techniques \citep{Kugel2023} to reproduce the observed $z=0$ galaxy mass function and the gas fractions of low redshift galaxy clusters. For an in-depth discussion of the galaxy formation model, we direct our readers to~\cite{Schaye:2023jqv}. Our analysis utilizes simulation boxes with resolutions of $900^3$ and $1800^3$ particles (i.e. simulations \texttt{L1\_m10} and \texttt{L1\_m9} introduced by~\cite{Schaye:2023jqv}). 

The evolution of each simulation's cosmological volume (from $z=15$ to $z=0$ ) is stored at 79 unequally spaced redshifts (``snapshots''). There are 61~snapshots between $z=3$ to $z=0$ (with redshift steps of $0.05$) and 8~snapshots are between $z=5$ and $z=3.25$;  the remaining 9~snapshots are spread between $z=5.5$ to $z=15$.

\subsection{Preparation of the light cone data}

\begin{figure}
    \centering
    \includegraphics[width=\linewidth]{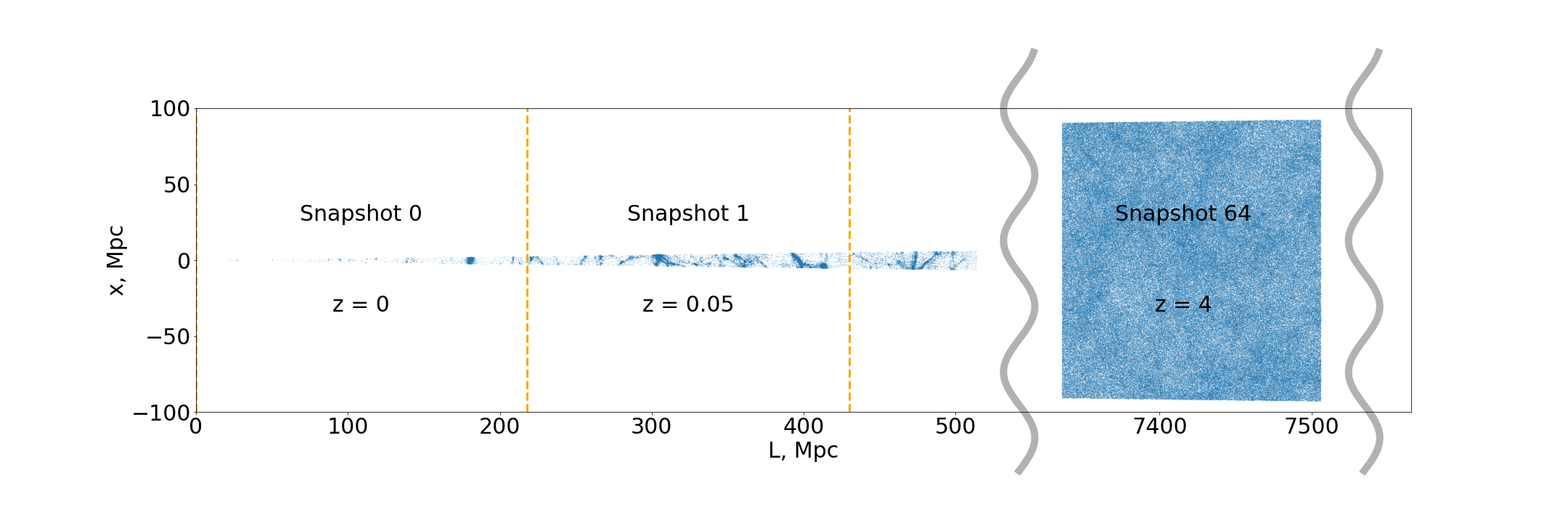}
    \caption{Visualization of the light cone construction process. The blue color encodes the electron number density within the light cone as a function of comoving distance~$L$. The orange dashed lines depict the borders between snapshots from which the data was taken.}
    \label{fig:Lightcones_sections_line}
\end{figure}

The extended cosmological volume of FLAMINGO creates an ideal basis for constructing light cones.  As shown in~\cite{CarlsonReshapeCosmoVol}, the periodic boundary conditions allow us to re-shape the cosmological volume into more elongated shapes along a particular axis. This technique allows for the remapping of FLAMINGO's already vast snapshot volume into a structure that progressively increases in redshift, approaching the next saved snapshot, thereby creating a continuous volume across a certain redshift span. Data for different sections of the light cone, as depicted in Fig.\ref{fig:Lightcones_sections_line}, is sourced from the corresponding snapshots. 
 We use a square pyramid as the light cone base shape. The cone itself has an aperture angle of $\alpha=4.5\cdot10^{-2}$~rad and spans from $z=0$ to $z=6$, corresponding a cross section of $(380\text{ cMpc})^2$ at $z=6$. We created three light cones for each of the used resolutions to mitigate random fluctuations. The light cones' observer positions and direction vectors are chosen randomly. We use the python package \textsc{SWIFTsimIO}~\cite{Borrow2020} to optimize and speed up the reading and handling of simulation data.

\subsection{Data projection and line-of-sight sampling}

This section elaborates on the method used to project data, particularly the electron number density along individual lines-of-sight (LOS). Each LOS is segmented into pixels, each $85~\rm{ckpc}$ in size. Within the FLAMINGO simulations, every gas particle is tagged with its electron number density. We calculate their contributions to each pixel using the following algorithm:
\begin{enumerate}
    \item For a given pixel we find all particles in the simulation, such that the distance from the particle to the center of the pixel is within the  smoothing length of a gas particle;
    \item For each selected particle we calculate its contribution to the given pixel using the smoothed particle hydrodynamics (SPH) kernel~\cite{BSpline}:
        \begin{align}
            K(q) = 
                    \begin{cases}
                    2\times\frac{8}{\pi}\times(1-q)^3, & \quad \frac12 < q < 1 \\[10pt]
                    \frac{8}{\pi}\times (1 - 6q^2 + 6q^3), & \quad q < \frac12
                \end{cases}  .
        \end{align}
        Here, $q$ is the distance between points in units of the SPH smoothing length.
    \item After summing up the contributions of different particles to the given pixel, we rescale the result as
    \begin{equation}
        n_e^{\text{final}} = n_e^{\text{sum}} \frac{(1+z_{\text{pixel}})^3}{(1+z_{\text{snap}})^3},
    \end{equation}
    where $z_{\text{pixel}}$ and $z_{\text{snap}}$ are the redshift of the specific pixel and the redshift of the snapshot from which we take the data respectively. This correction allows us to have a smoother evolution of the electron number density along the LOS~\cite{Garcia:2020qrp}.
\end{enumerate}

\begin{figure}
    \centering
    \includegraphics[width=.9\linewidth]{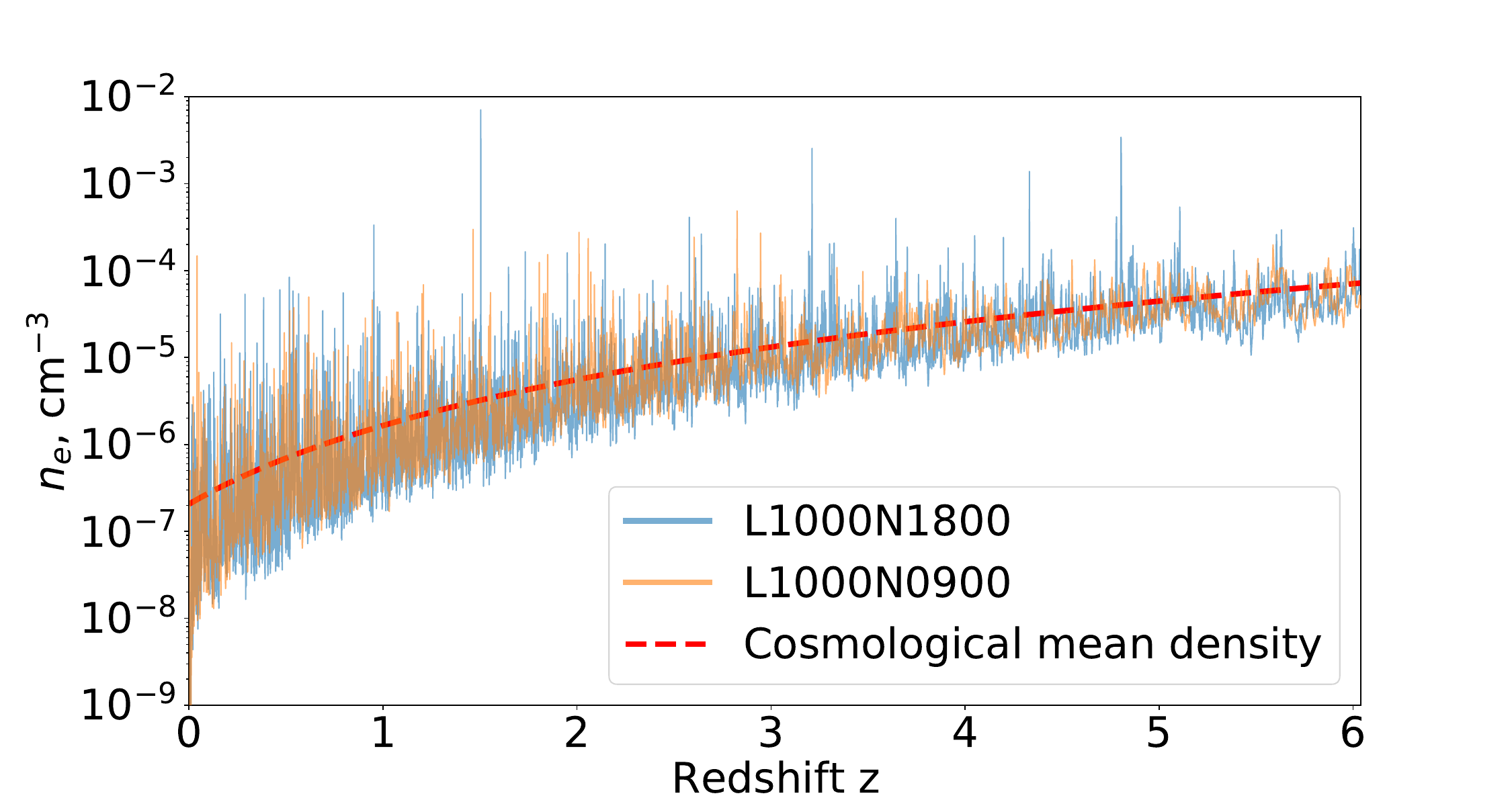}
    \caption{Electron number density along a random line-of-sight in the light cone volume for the low-resolution simulation (blue line) and high-resolution simulation (orange line). The dashed red line indicates the theoretical cosmological average electron number density.}
    \label{fig:neaverage}
\end{figure}

An example of the electron number density along a random line of sight for different simulation resolutions is presented in Fig.~\ref{fig:neaverage}. Upon comparing the low- and high-resolution simulation results, it is evident that the higher-resolution simulation exhibits a greater number of peaks with increased electron number density.\footnote{For example, for $n_{e,\text{res}} = 10^{-3}\text{ cm}^{-3}$ there are 104739 points of resonance along all lines of sight for lower resolution of the simulation, while for higher resolution there are 153857 points of the resonance for the same number lines of sight.} This observation aligns with expectations, as a higher-resolution simulation can resolve smaller and, hence, denser structures.

\subsection{Predictions for the conversion probability}

In this study, we aim to compare the anisotropies from dark photon conversion with the CMB power spectrum observed by the Planck satellite. Consequently, we align the pixel size in our simulated sky with the angular resolution of Planck, which corresponds to squares measuring $1.5\cdot 10^{-3}$ radians. This results in an image of $30\times 30$ pixels in our light cone. Nonetheless, it is possible for the conversion probability to exhibit fluctuations within a single pixel. To address this variability, we refine our approach by incorporating 9~LOS within each pixel. We calculate the conversion probability along each LOS using~\eqref{eq:ptot}. Subsequently, we take the average of the conversion probabilities within each pixel.

\begin{figure}
    \centering
    \includegraphics[width=\linewidth]{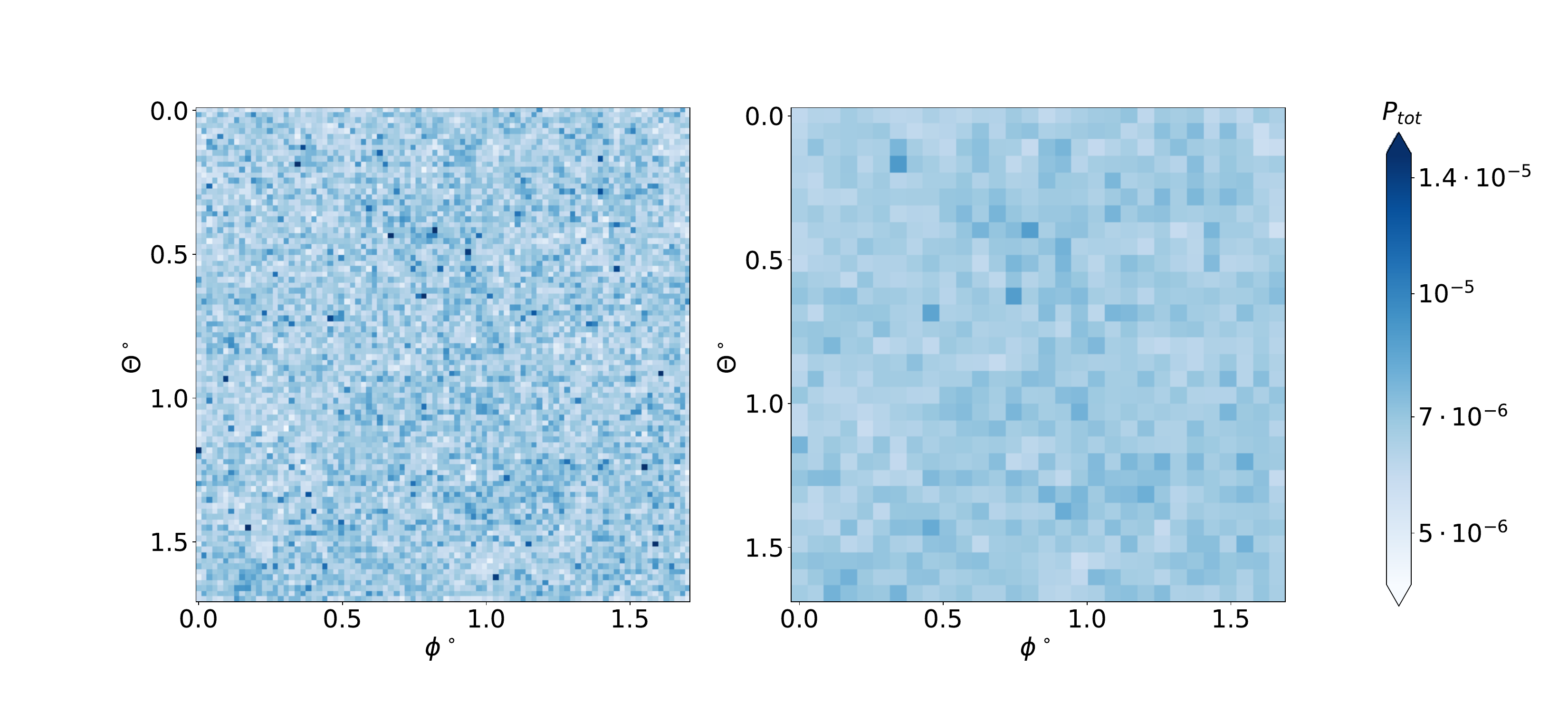}
    \caption{Two-dimensional visualization of the photon-to-dark-photon resonant conversion probability along the LOS within the simulated light cone. The left panel presents a grid configuration of $90 \times 90$ LOS, where each pixel corresponds to an angular size of $0.5 \times 10^{-3}$~rad, which is three times finer than the resolution of the Planck satellite~\cite{Planck:2018bsf}. The right panel displays the conversion probability, averaged over batches of $3 \times 3$ LOS, such that each pixel is at Planck's resolution. The map of conversion probabilities is shown for a dark photon mass $m_{A'} = 1.2\times 10^{-13}$~eV and kinetic mixing~$\epsilon = 10^{-6}$.}
    \label{fig:conv_prob}
\end{figure}

The probability of resonant conversion depends on the mass of the dark photon. 
We calculate maps of conversion probability within the mass range $3.7\times 10^{-15}<m_{A'}/\text{eV}< 3.7\times 10^{-12}$. This corresponds to resonant electron number densities $10^{-8}<n_e/\text{cm}^{-3}<10^{-2}$. These values are covered by simulation data within our redshift range $0 \leq z \leq 6$ (see Fig.~\ref{fig:neaverage}).
An example map of conversion probability for $m_{A'} = 1.2\times 10^{-13}$~eV (corresponding to $n_e=10^{-5}\text{ cm}^{-3}$) is shown in Fig.~\ref{fig:conv_prob}.

\section{Comparison with CMB data and results}
\label{sec:results}

In this section, we place limits on the dark photon model by comparing the light cone predictions obtained from simulation data with existing CMB temperature anisotropy data. For this, we map the energy-dependent flux decrement predicted by the conversion probability to temperature fluctuations of the CMB. 

A major difficulty in applying such predictions to actual data is, however, the complex post-processing pipeline of CMB experiments designed to remove foregrounds. Such a procedure could in fact subtract a contribution 
from dark photon conversion from the reported final power spectrum. To deal with this issue, we chose to instead use {\it raw}, foreground-unsubtracted CMB data. Demanding that the conversion-induced power does not overshoot this data then results in the most conservative and map-making-independent constraint.

At a given photon frequency $\omega$, a change of  flux $\Delta F(\omega)$ leads to a temperature shift $\Delta T(\omega)$ as follows
\begin{equation}
    \Delta F(\omega) = - b_\omega(\omega, T) P_{\text{tot}}(\omega) = \frac{d b_{\omega}}{dT} (\omega, T)\Delta T(\omega).
\end{equation}
Here, $b_\omega (\omega, T)$ is the initial CMB photon flux for which we take the Planck blackbody spectrum at temperature~$T$. We then obtain,
\begin{equation}
    \Delta T(\omega) = - P_{\text{tot}} \frac{b_{\omega}}{db_{\omega}/dT} = - \frac{(1 - e^{-\omega/T}) P_{\text{tot}}(\omega) T^2}{\omega}.
    \label{eq:DeltaT_at_omega}
\end{equation}
Taking into account that $P_{\text{tot}}(\omega) \propto 1/\omega$, we conclude that $\Delta T(\omega)$ is a monotonically decreasing function with energy. Thus, we expect the strongest additional anisotropies from dark photons at the lowest accessible photon frequencies. 

\begin{figure}
    \centering
    \includegraphics[width=\linewidth]{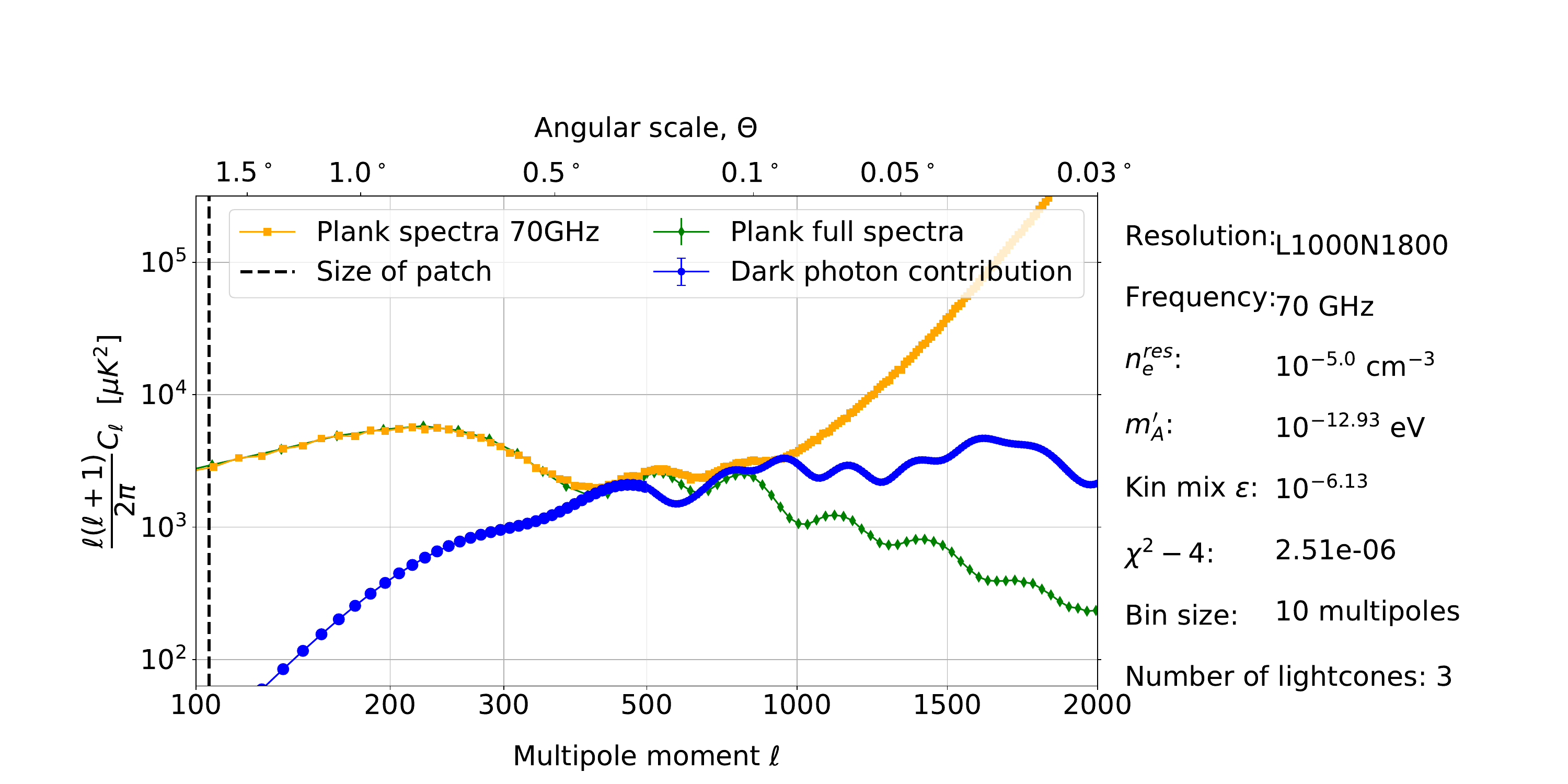}
    \caption{Comparison of the temperature power spectrum induced by resonant conversion (blue line) with the raw observed CMB temperature power spectrum measured by the Planck satellite at 70~GHz with the galaxy geometrically masked (orange line). The green line shows the Planck-reported temperature power spectrum after foreground-cleaning~\cite{Planck:2018bsf}. The vertical black dashed line shows a minimal multipole moment resolved by our simulation data. The power spectrum from the conversion is obtained for  $m_{A'}=1.2\times 10^{-13}$~eV and $\epsilon = 7.4\times 10^{-7}$, which is a point at the boundary of our exclusion limit.
    } 
    \label{fig:deltaT}
\end{figure}

In this work, we use Planck measurements at waveband $70$~GHz~\cite{Planck:2018bsf}.\footnote{There are two Planck wavebands of lower frequency, $30$ and $44$~GHz. Measurements at these frequencies are, however, more polluted by foregrounds.} As outlined above, to be sure that the foreground-removal procedure does not influence our result, we calculate the temperature power spectrum ourselves, using the raw image data from the Planck legacy archive. For this, we take the sky map\footnote{In the HEALPix map, the sphere is divided into 12 equal-area regions, which corresponds to NSIDE$=1$. The resolution increases by subdividing each region into NSIDE$^2$ pixels of equal area.} with $\text{NSIDE} = 1024$ and apply a simple mask that excludes data with galactic latitude $|\theta|< 30\degree$ to largely remove the Milky Way from the raw data geometrically. We then calculate the temperature power spectrum using the algorithm \texttt{Anafast} from the python library \texttt{healpy}~\cite{Zonca2019} which is based on the \texttt{HEALPix} scheme~\cite{2005ApJ...622..759G}. The obtained power spectrum is then averaged over 10 multipole moments to decrease noise and divided by the square of the beam window function as a first-order correction of the beam.\footnote{This correction is essential to mitigate beam smoothing and to accurately infer the angular power spectrum at scales smaller than the telescope's resolution~\cite{Planck:2019nip}.} 

The result is shown by the orange line in Fig.~\ref{fig:deltaT}. We see that for $\ell \lesssim 300$, the calculated power spectrum is very close to the final reported one from Planck. At larger $\ell$, we observe significant pollution from different foregrounds, which are not masked in our data.

To estimate the contribution from dark photon conversion, we used the HEALPix framework with the same resolution parameter. In our approach, we mask all pixels outside a specified central region of $30 \times 30$ pixels, effectively focusing on the area of primary interest. We then insert an array representing $\Delta T$ values, calculated using Eq.~\eqref{eq:DeltaT_at_omega}, into the $30 \times 30$ region within the HEALPix map. Using the obtained masked map, we compute the power spectra with the same \texttt{Anafast} algorithm, concentrating on the ranges of multipole moments that align with the sizes of both the individual pixels and the entire selected patch and average the result for three realizations of the light cone. Finally, we average the power spectra over 10 multipole moments to obtain a smoother estimation. An example power spectrum from dark photon conversion for the higher-resolution simulation is shown by the blue line in Fig.~\ref{fig:deltaT}. We find that lower and higher resolution FLAMINGO simulations yield agreeing results, which implies that numerical convergence has been reached; a study of resolution effects is presented in App.~\ref{app:resolution-study}. We observe that the dark photon-induced power spectrum grows with~$\ell$. This behavior is expected as the conversion is correlated with the Universe's large-scale structure, which increases towards smaller angular sizes.

To obtain the exclusion limits, we compare the temperature power spectrum $D_\ell^{A'}$ from dark photon conversion with the calculated raw CMB spectrum from the Planck data at 70~GHz ($D_\ell^{Planck}$). Constraints on the kinetic mixing parameter $\epsilon$ are then inferred from a $\chi^2$ test. We require that $\Delta\chi^2 < 4$,  corresponding to a $2\sigma$ ($\sim95\%$) confidence level,
\begin{equation}
   \Delta\chi^2=\sum_i\dfrac{(D_{\ell,i}^{A'}-D_{\ell,i}^{Planck})^2}{\left(\Delta D_{\ell,i}^{A'}\right)^2},
\end{equation}
and the sum is only taken over bins $i$ for which the prediction overshoots the data, $D_\ell^{A'} > D_\ell^{Planck}$.

The approach described above gives us very conservative limits on dark photon mixing. To estimate the potential reach of this method with analysis of full Planck data, we take the final Planck TT power spectrum and fit the CMB spectrum with a prediction with dark photons changing cosmological parameters ($H_0$, $\Omega_{\text{DM}}$, $\Omega_b$, $A_s$, and $n_s$). We refit cosmological parameters for different $\epsilon$ values and find an optimistic estimate of the $2\sigma$ constraint on dark photon mixing when the quality of fit changes by $\Delta \chi^2 = 4$.

\begin{figure}
    \centering
    \includegraphics[width=\linewidth]{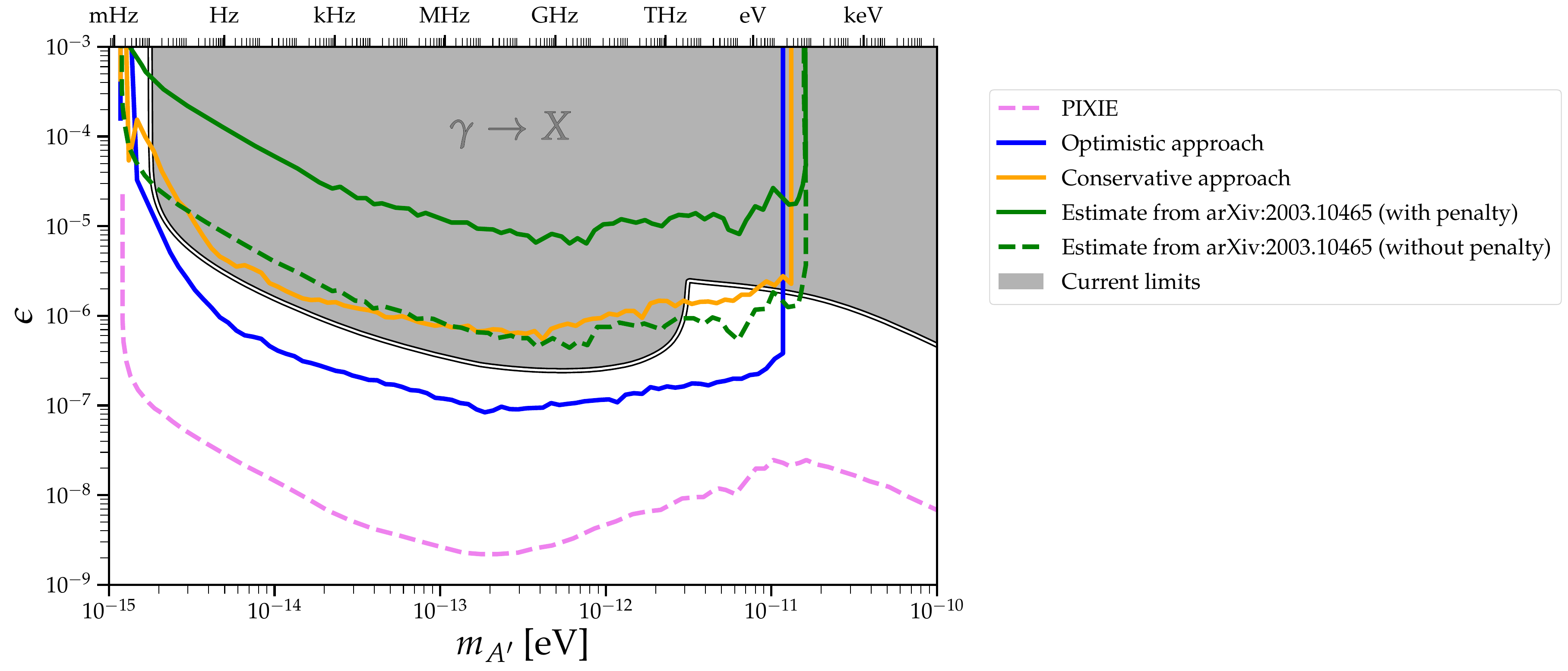}
    \caption{Conservative and optimistic constraints on the dark photon mixing angle from CMB anisotropies for different dark photon masses (orange line and blue line correspondingly). Estimates of Planck sensitivity using the full-sky variance of temperature fluctuations are shown with green solid and dashed lines~\cite{Garcia:2020qrp}. Previous limits in this mass region are cumulatively shown by the grey area~\cite{Caputo:2020bdy, Garcia:2020qrp} and obtained with the help of~\cite{Caputo:2021eaa}. The estimate on the PIXIE-type sensitivity (pink dashed line) is taken from~\cite{Garcia:2020qrp}.}
    \label{fig:epsilon}
\end{figure}

The constraints on dark photon kinetic mixing $\epsilon$ are shown in Fig.~\ref{fig:epsilon} by orange line (conservative limit) and blue line (estimate from full Planck data). We see that the conservative approach gives similar constraints as an estimate from~\cite{Garcia:2020qrp}, which is slightly below the current constraints from spectral distortions~\cite{Garcia:2020qrp,Caputo:2020bdy} with a small improvement in the high-mass range. However, usage of the full Planck dataset should allow to improve current limits on dark photon mixing angle in the mass interval $3\times 10^{-15}\text{ eV} < m_{A'} < 3\times 10^{-12}\text{ eV}$ from a factor of a few up to one order of magnitude (for the high mass range).
For smaller dark photon masses, the conversion probability is susceptible to the properties of the local volume, while the resolution of the used simulation is insufficient to map regions with larger~$n_e$, see App.~\ref{app:resolution-study} for details.

\section{Summary and conclusions}
\label{sec:conclusions}

In this work, we derive the power spectrum of anisotropies imprinted on CMB photons by their conversion into dark photons in the post-reionization Universe. To accurately account for the distribution of electron number density---which acts as the agent in resonant $A\to A'$ conversion, we utilize the newly available large-volume FLAMINGO simulation suite~\cite{Schaye:2023jqv}. Using its data at various reshifts between $0\leq z \leq 6$, we construct light cones that allow us to simulate photon propagation over cosmological distances. 
From there, we compute the total probability that a CMB photon leaks into the dark sector, and subsequently the resulting effective CMB temperature fluctuation and associated power spectrum. 

Using raw Planck data from the $70$~GHz waveband we put conservative limits on the dark photon model, which are slightly weaker than previous limits from spectral distortions. We also showed that by using full Planck data one can achieve much better sensitivity improving our result by approximately one order of magnitude. 

To perform an analysis of full Planck data, several improvements compared to this work should be made. First, one may use observational data from several wavebands and leverage the frequency dependence of the conversion probability as resonant conversion results in spectral distortions of a specific shape. A dedicated algorithm to search for this specified type of signal should increase sensitivity. Second, we use raw CMB data to protect ourselves from biases that could emerge from the signal subtraction introduced by the Planck collaborations' analysis pipeline. However, one can certainly define a strategy to foreground-clean and combine data from different wavebands such that it does not remove any additional signal from conversion. We expect that those improvements will strengthen the limit by at least a factor of a few. Such investigations are left for future work.

Finally, it is worth mentioning, that the best method to achieve a breakthrough in this field is to develop a next generation of spectral distortion experiments such as PIXIE~\cite{Kogut:2011xw} (see Fig.~\ref{fig:epsilon}). Preliminary estimates made in~\cite{Garcia:2020qrp} show that PIXIE data should allow us to strengthen our sensitivity to dark photons by a few orders of magnitude. 

\begin{figure}[tbp]
    \includegraphics[width=\linewidth]{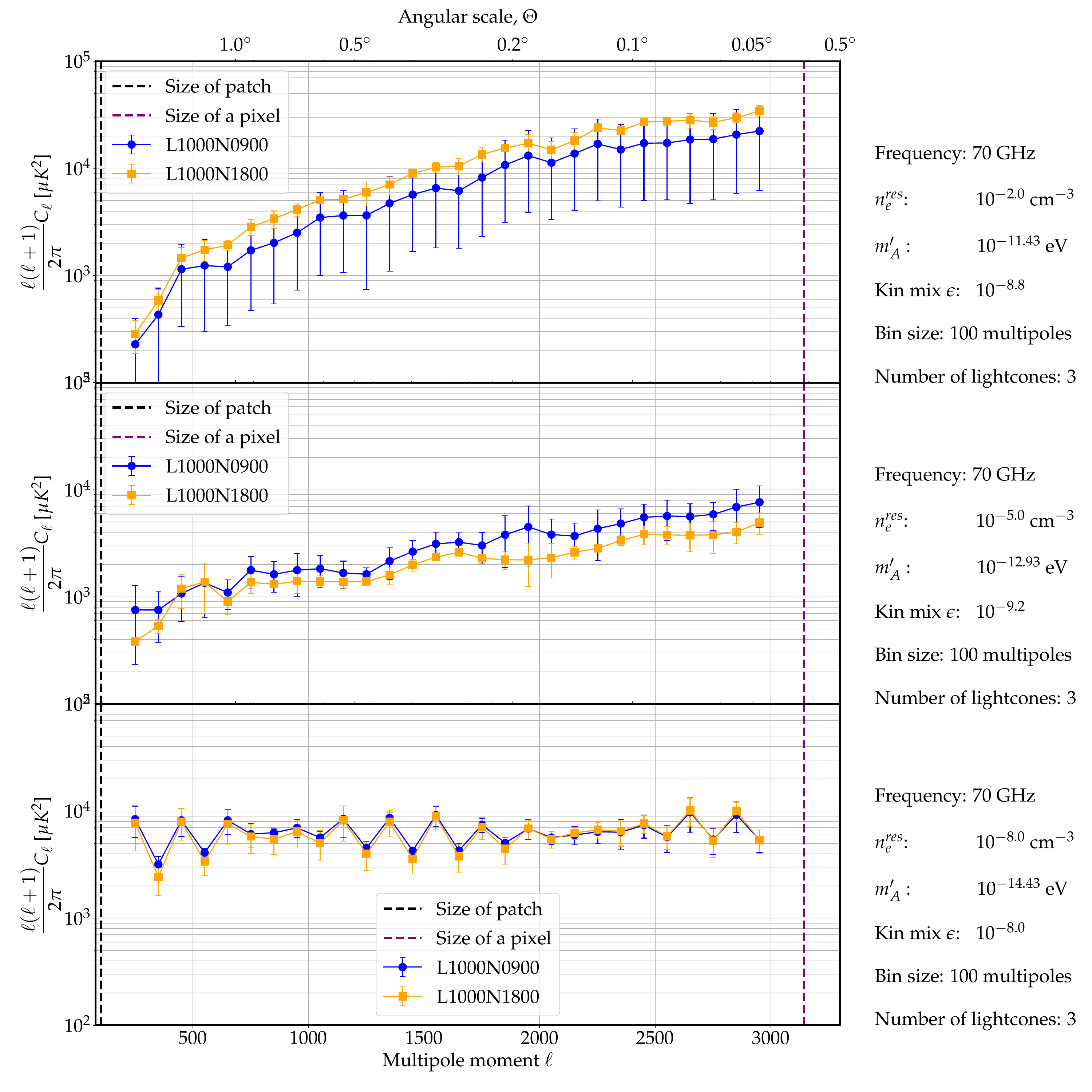}
    \caption{Comparison of calculated temperature power spectra for different simulation resolutions (L1000N0900 and L1000N1800) for three different resonant electron number densities: $n_e^{\text{res}} = 10^{-2}\text{ cm}^{-3}$ (upper panel), $n_e^{\text{res}} = 10^{-5}\text{ cm}^{-3}$ (middle panel), and $n_e^{\text{res}} = 10^{-8}\text{ cm}^{-3}$ (bottom panel).
    } 
    \label{fig:resolution_DeltaT}
\end{figure}

\begin{figure}[h!]
    \centering
    \includegraphics[width=1.1\linewidth]{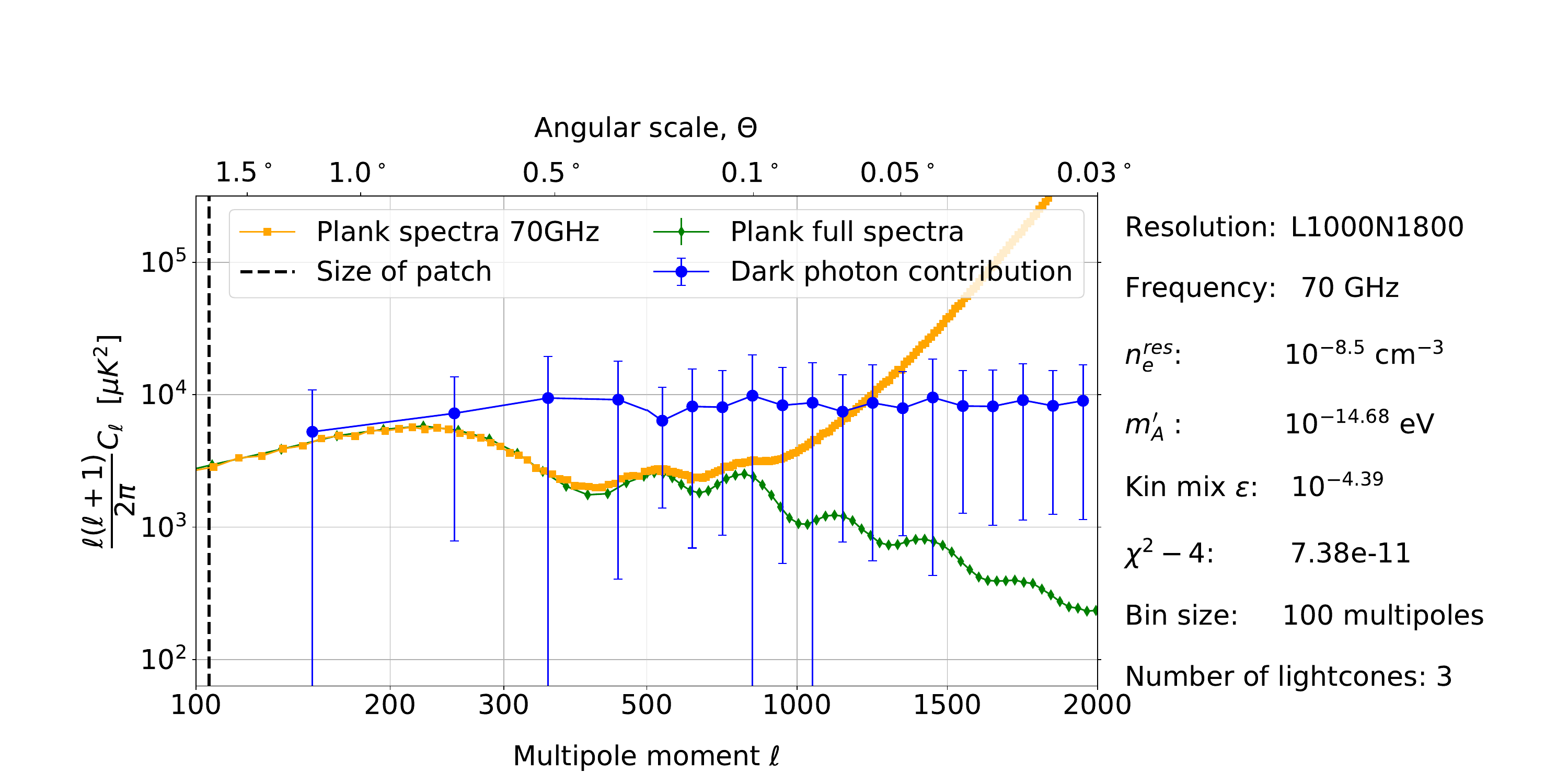}
    \caption{The same as Fig.~\ref{fig:deltaT}, but for a dark photon mass $m_{A'} = 3.7\times 10^{-15}$, which corresponds to a resonant electron number density  of $n_e^{\text{res}} = 3.2\times 10^{-9}\text{ cm}^{-3}$.}
    \label{fig:DeltaT_ne-8.5}
\end{figure}

\begin{figure}[t]
    \centering
    \includegraphics[width=.9\linewidth]{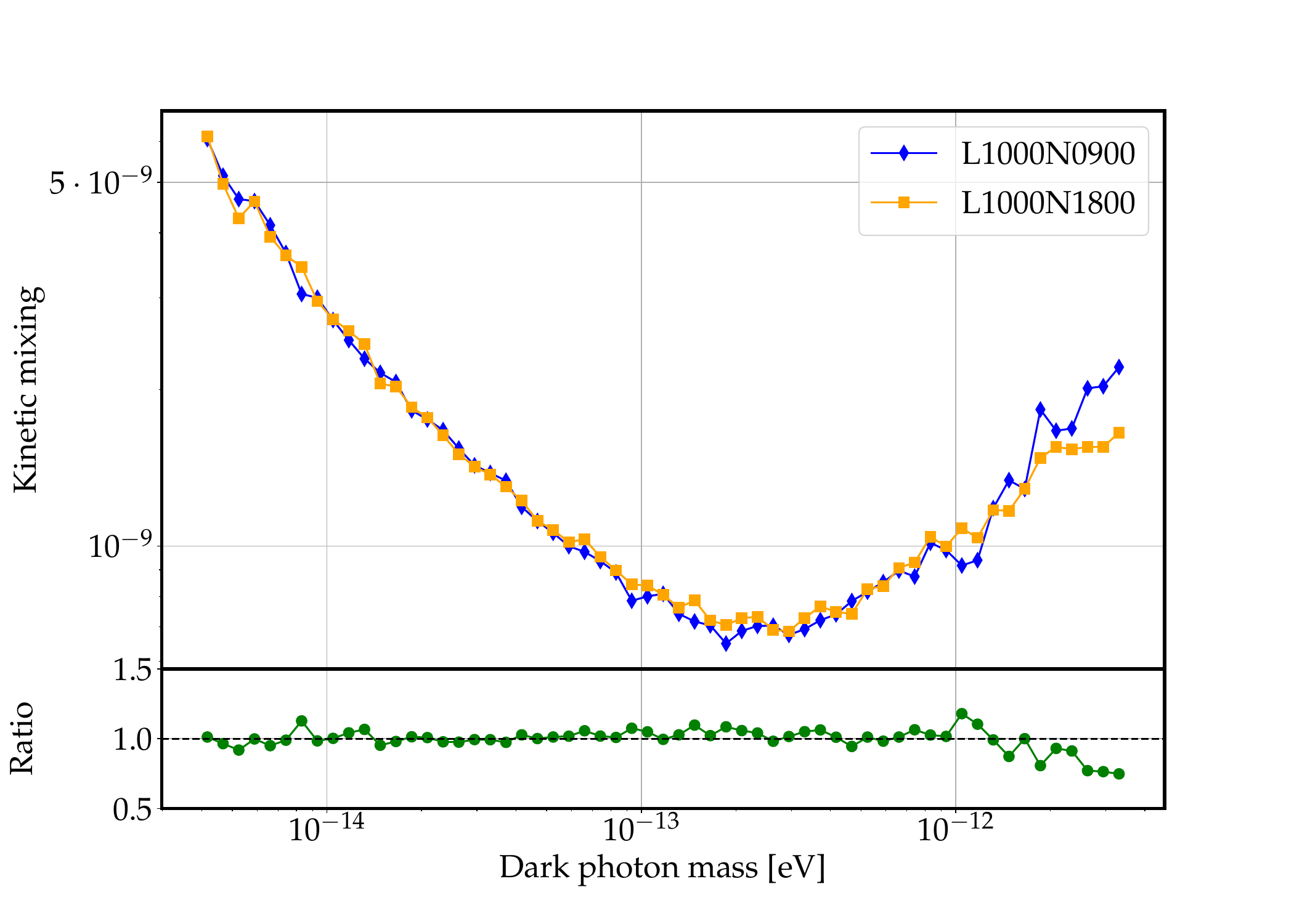}
    \caption{Comparison of constraints on the kinetic mixing~$\epsilon$ for the two different resolution FLAMINGO simulations (blue line for lower resolution and orange for higher one). The bottom panel shows the ratio of constraints for higher and lower resolutions.}
    \label{fig:epsilon_resolution_study}
\end{figure}

\section*{Acknowledgements} We would like to thank Albert Stebbins and Julien Lesgourgues for helpful discussions.  We acknowledge support from {\sl Fermi Research Alliance, LLC} under Contract No. DE-AC02-07CH11359 with the U.S. Department of Energy, Office of High Energy Physics.
Based on observations obtained with Planck (\url{http://www.esa.int/Planck}), an ESA science mission with instruments and contributions directly funded by ESA Member States, NASA, and Canada. Some of the results in this paper have been derived using the healpy and HEALPix package~\cite{Górski_2005}. This work used the DiRAC@Durham facility managed by the Institute for Computational Cosmology on behalf of the STFC DiRAC HPC Facility (www.dirac.ac.uk). The equipment was funded by BEIS capital funding via STFC capital grants ST/P002293/1, ST/R002371/1 and ST/S002502/1, Durham University and STFC operations grant ST/R000832/1. DiRAC is part of the National e-Infrastructure. This work was supported by the Research Network Quantum Aspects of Spacetime (TURIS). This research was partially supported by an iBOF project [iBOF/23/074] of the Special Research Fund (Bijzonder Onderzoeksfonds, BOF) of the Flemish government.

\appendix

\section{Resolution study}
\label{app:resolution-study}

The FLAMINGO suite includes cosmic simulations with different resolutions and sizes: L1000N0900, L1000N1800, L1000N3600, and L2800N5040. Due to time and computational limits, we limit our resolution study to the first two boxes with the smallest amount of particles. We repeat the whole procedure, starting from the creation of light cones to calculating power spectra similarly, for two resolutions. The obtained spectra for three different resonant electron number densities (minimum, median, and maximum) are shown in Fig.~\ref{fig:resolution_DeltaT}. As one can see, the spectra for different simulation resolutions agree with each other within the error bars. For the high value $n_e^{\text{res}}=10^{-2}~\text{cm}^{-3}$ it can be noticed that the spectra of L1000N1800 are systematically elevated. This is expected since the higher resolution simulation allows resolving denser structures (see Fig.~\ref{fig:neaverage} as an example), that are associated with high electron number densities. For the same reason, there is a mildly smaller amount of pixels with average $n_e^{\text{res}}=10^{-5}~\text{cm}^{-3}$. Therefore, the spectrum of L1000N0900 is higher than for the higher-resolution simulation. For low resonant number densities, we stop our estimates at $n_e^{\text{res}}=10^{-8}~\text{cm}^{-3}$, because for even lower densities, the resulting power spectrum starts to fluctuate significantly, see Fig.~\ref{fig:DeltaT_ne-8.5}. This is because, for such low electron number densities, the probability of the conversion depends strongly on the properties of the local volume near the beginning of the light cone (see Fig.~\ref{fig:neaverage}).

The inferred upper limits on the kinetic mixing parameter~$\epsilon$ for the two different resolutions are shown in Fig.~\ref{fig:epsilon_resolution_study}. This plot ascertains, that increasing the resolution only slightly strengthens the constraint and only for high electron number densities (large dark photon masses). This means that our results based on L1000N1800 simulation are a conservative bound, as any increase in resolution will lead to even stronger constraints on the kinetic mixing in the high-mass dark photon region.

\bibliographystyle{JHEP}
\bibliography{refs.bib}

\end{document}